\documentclass[12pt]{iopart}

\usepackage[square,sort&compress,comma,numbers]{natbib}
\usepackage{graphicx}
\usepackage{float, lineno, amssymb, hyperref, enumerate}
\usepackage[utf8]{inputenc}

\begin{document}
\title{Epidemics in networks: A master equation approach}
\author{M. Cotacallapa$^{1,\dagger}$, M. O. Hase$^{1,\ddagger}$}

\address{$^1$Escola de Artes, Ciências e Humanidades, Universidade de São Paulo, Avenida Arlindo Béttio 1000, 03828-000, São Paulo, SP, Brazil}
\ead{$^\dagger$moshe@usp.br, $^\ddagger$mhase@usp.br}

\begin{abstract}
A problem closely related to epidemiology, where a subgraph of \textquotedblleft infected\textquotedblright links defined inside a larger network is investigated. This subgraph is generated from the underlying network by a random variable, which decides whether a link is able to propagate a disease/information. The relaxation timescale of this random variable is examined in both annealed and quenched limits, and the effectiveness of propagation of disease/information is analyzed. The dynamics of the model is governed by a master equation and two types of underlying network is considered: one is scale-free and the other has exponential degree distribution. We have shown that the relaxation timescale of the contagion variable has major influence in the topology of the subgraph of \textquotedblleft infected\textquotedblright links, which determines the efficiency of spreading of disease/information over the network.
\end{abstract}


\vspace{2pc}

\submitto{\jpa}

\maketitle

%% Content
\section{Introduction}
\label{introduction}

The analysis of structure and dynamics of many \textquotedblleft real\textquotedblright networks, like the WWW \cite{Barabasi2000}, Internet \cite{Faloutsos:1999:PRI:316194.316229}, and many biological \cite{Jeong2000} and social \cite{Barabasi2002} ones, inspired many works in the last years. Nowadays, from the massive data available, one can recognize a heterogeneous structure in many of these (complex) networks, usually following a power-law degree distribution. From the standpoint of their representation as graphs, the links between vertices show the interconnections between different nodes, and the degree distribution is a relevant aspect of transmission of any kind of information over the network. The spreading of information between vertices is closely related to the classical problem of percolation \cite{doi:10.1021/ja01856a061,PSP:2048852}, which was first examined on Bravais lattices or other regular structures like Cayley trees \cite{Stauffer1994}.

An important question that arises is associated to the problem of epidemiology on complex networks. First of all, in the study of epidemiology, the disease should propagate between people that are interconnected in a fashion which resembles a complex network, and not a regular lattice. Moreover, since critical phenomena on complex networks display a richer behaviour than classical statistical models on regular lattices (for instance, the first results of ferromagnetic Ising model on Barab\'asi-Albert model can be found in \cite{Aleksiejuk2002,Bianconi2002}; see also a review \cite{RevModPhys.80.1275} for critical phenomena on complex networks), one expects richer results when dealing with problems defined on complex networks.

In the literature, it is known that a SIS (susceptible-infective-susceptible) epidemic model on a scale-free network displays persistence, and no positive critical value $\lambda_{c}$, below which the epidemy is confined on a finite fraction of the infinite network, exists \cite{Pastor-Satorras2001,Pastor-Satorras2001b,May2001,Boguna2003}. This problem is closely related to the problem of robustness of a network, where the connectivity of a graph is examined as vertices are removed, and this problem can be mapped on a percolation problem \cite{Albert2000,Cohen2000,Cohen2001,Callaway2000}. It is known, since then, that scale-free networks are robust against failures (removal of random vertices), but weak against attacks (removal of targeted nodes).

In this work, the problem of propagation of a disease or information will be revisited in the framework of a master equation combined with a disordered variable that mimics the propagation rate. This random variable will be analyzed in two extreme regimes: annealed and quenched. These terminology are commonly present in disordered systems (in particular, spin glass theory \cite{mezard1987spin}), and the problem of epidemiology (or propagation of informations over the graph) on a complex network will be viewed as a disordered problem in this work.

Given a network, an edge that links two vertices represents an interactions between them; nevertheless, this connection is a potential channel of communication, but does not imply, necessarily, that these two nodes share or transmit an information or a disease. In this work, we characterize the subgraph of this network that effectively propagates an information/disease, and see how this subgraph relates to the underlying network. In this sense, our work differs from traditional studies of epidemiology on complex networks commented in the previous paragraph, since we are not investigating the emergence of a giant component of infected nodes and establishing a threshold in the infection rate as a percolation problem. We did not consider the traditional dynamical process of disease/information spreading, but we characterized the subgraph of the network composed by nodes that can effectively communicate each other to transmit disease/information when it appears. Therefore, we are interested in the statistics of degrees of the nodes that can participate in the process of propagation of information/disease (if it appears) in the network. This is a measure of how an information/disease can propagate over the network, and we have introduced the \textquotedblleft reduced degree distribution\textquotedblright (see section \ref{models}) to quantify it.

Moreover, we have examined the graph which effectively transmits information/disease in two cases, as mentioned before. When an edge is recognized as a true transmitter of information/disease between two vertices, we considered a case where this connection remains so and does not close the flow of information/disease. This is the case where the random variable that dictates if a link is \textquotedblleft functional\textquotedblright or not is a quenched variable. We have also considered the other extreme situation, where the change of the state of links (transmit/not transmit) is rapid (the annealed case).

The layout of this work is as follows: in section \ref{models} we present the basic setup for the models that are analyzed in two different regimes of contagion process at sections \ref{fixed} and \ref{contagion}. The results are discussed in the last section.
\section{Epidemic models}
\label{models}

In this section, we give a summary of the models that will be considered in this work. To develop the epidemic models, two questions are made in the construction of the network: \textquotedblleft which node will we link?\textquotedblright and \textquotedblleft will we be infected by this node?\textquotedblright. In the present case, the first question is answered by the rules of the network dynamics, and the second question is where this paper introduces the contagion process. 

The main setup is a growing network where a single vertex is added at each time step to the graph, and this new incoming vertex links with just one single old pre-existing one with probability $\Pi(k, t)$ at time $t$, where $k$ is the degree of the old node that receives the link from the new one. Once this connection is made, we draw a random variable $r_{t}$ to decide if this link can transmit disease (or information) or not. This two-steps dynamical process can be described by the discrete time master equation
\begin{eqnarray} \label{eq:master-random-network}
p_{r}(k,s,t+1) = r_{t}\Pi(k-1, t)p_{r}(k-1,s,t) + \left[1-\Pi(k, t)\right]p_{r}(k,s,t)\,,
\end{eqnarray}
where $p_{r}(k, s, t)$ stands for a mesure of the site $s$ having degree $k$ at time $t$ in the subgraph of vertices that can propagate information/disease. Note that the label $s$ of a vertex stands also for the instant when this vertex joined the network. The boundary condition, associated to this quantity for a new incoming vertex is, then, $p_{r}(k,s,t=s)=\delta_{k,1}$. For the random variable $r_{t}$, which mimics the contagion process, we choose a Bernoulli distribution,
\begin{eqnarray}
\mathcal{P}(r_{t}) = \left(1-c\right)\delta_{r_{t},0} + c\delta_{r_{t},1}\,,
\label{eq:bernoulli}
\end{eqnarray}
where $\delta$ is the Kronecker symbol and $c\in[0,1]\subset\mathbb{R}$ controls the contagion rate. For notational convenience, we adopt two vertices that have two connections between them at time $t=2$ as the initial condition, but this particular choice is not expected to have serious impact on the properties we are interested after a long time. In this setup, the total degree of the graph at time $t$ is simply $2t$.

We will now discuss some dynamics chosen for the growing network above in this work. In the first one, at each unit of time a new node is added and randomly links to another old node chosen with equal probability. In other words, at time $t$, one has $\Pi(k, t)=\frac{1}{t}$. After that, it decides whether or not to infect the old node. The second dynamics is based on a preferential attachment scheme, and $\Pi(k, t)\propto k$ \cite{1999Sci...286..509B}: vertices that have higher degrees (infectious or not) are more likely to receive new connections.

Furthermore, we will investigate the role of the contagion process, represented by the set of random variable $\{r_{s}\}$. Let us examine the equation (\ref{eq:master-random-network}), which provides information on the subgraph of vertices that can propagate informations/disease. A vertex $s$ of this subgraph can have degree $k$ at time $t+1$ if:

\begin{enumerate}
  \item The vertex $s$ of this subgraph has $k-1$ links at time $t$, and receives a link from the new incoming vertex with probability $\Pi(k-1,t)$. This is the term $\Pi(k-1,t)p_{r}(k-1,s,t)$. Then, two situations should be considered:
  \begin{enumerate}
    \item This new link allows the propagation of information/disease, $r_{t}=1$. This implies the contribution $r_{t}\Pi(k-1,t)p_{r}(k-1,s,t)$.
    \item This new link does not allow the propagation of information/disease, $r_{t}=0$. This implies the contribution $r_{t}\Pi(k-1,t)p_{r}(k-1,s,t)=0$ to $p_{r}(k,s,t+1)$
  \end{enumerate}
  \item The vertex $s$ of this subgraph has already $k$ links at time $t$; therefore, it must not receive the connection from the new incoming vertex to keep degree $k$, and this happens with probability $1-\Pi(k,t)$.
\end{enumerate}

We will consider two limit cases of its relaxation time: the annealed and quenched ones. In the former, the contagion variable $r_{t}$ varies much faster than a typical \textquotedblleft observational time\textquotedblright of the whole process, and this fast fluctuation leads us to treat it as effectively having its mean value, $\langle r_{t}\rangle=c$, in respect to the distribution (\ref{eq:bernoulli}). In this annealed case, all the links transmit information/disease with probability $c$. In the other limit, the random variable $r_{t}$ is considered to be quenched. This means that during a typical \textquotedblleft observational time\textquotedblright, it remains constant (with $r_{t}=1$ or $r_{t}=0$) without any fluctuation. The analysis should then consider many realizations of the process, and one should evaluate $\langle p_{r}(k,s,t)\rangle$ in the end. It is important to stress that in the present work, vertices have the state \textquotedblleft transmits disease/information\textquotedblright (when it appears) and \textquotedblleft does not transmit disease/information\textquotedblright (even if appearing). Nodes may be connected, but not always they share or transmit informations (or diseases), and by \textquotedblleft contagion\textquotedblright between two vertices we mean that not only they are connected, but there is also a flow of information (disease) allowed between them. In this sense, this work is different from the previous ones treated in the literature \cite{Pastor-Satorras2001,Pastor-Satorras2001b,May2001,Boguna2003,Albert2000,Cohen2000,Cohen2001,Callaway2000}. It is worth to mention that, in the context of opinion dynamics, a dynamics with another timescale (time-dependent flipping rate of the voter model) was considered \cite{Baxter2011}.

Our goal is to examine the \textquotedblleft reduced degree distribution\textquotedblright $ \tilde{P}_{r}(k)$, which is defined as
\begin{eqnarray}\label{eq:Pred}
 \tilde{P}_{r}(k) = \lim_{t\rightarrow\infty}\frac{1}{t}\int_{0}^{t}ds\,p_{r}(k,s,t)\,.
\end{eqnarray}
Let us first consider the case without contagion variable to understand (\ref{eq:Pred}): if $p(k,s,t)$ is the probability of a vertex $s$ having $k$ links at time $t$, the time-dependent degree distribution $P(k,t)$ is the average value of $p(k,s,t)$ over the sites, $P(k,t)=\sum_{s=1}^{t}p(k,s,t)/t$, and the degree distribution is its stationary value, $P(k)=\lim_{t\rightarrow\infty}P(k,t)$. The \textquotedblleft reduced degree distribution\textquotedblright $\tilde P_{r}(k)$ above is inspired in this definition. This quantity, $\tilde P_{r}(k)$, is related to the degree distribution of the subgraph of vertices that allow propagation of information/disease, and we will analyze different systems to examine if it differs (and how it differs) from the original degree distribution of the underlying network. Nevertheless, it should be noted that the function $ \tilde{P}_{r}(k)$ is not normalized for $c\neq 1$ (therefore, it is not a probability distribution), because it counts the fraction of nodes that has transmitting links with respect to the whole network (and not to the vertices of the infected subgraph only). However, it gives a measure of the distance to the process where the contagion always happens (which is the case $r_{s}=1$ for any $1\leq s\leq t$, or $c=1$).

\section{Fixed contagion process}
\label{fixed}

As a preliminar analysis, we will consider the random variable $r_{s}$ ($1\leq s\leq t$), which decides the flow of infection on a edge, as being the same for every link and no changes will be made over this variable while the network is growing. In this section, therefore, we will denote this random variable as $r=r_{s}$. This crude approximation will be compared later to a more realistic situation where each link has an independent configuration of propagation. From this setup, two limit cases of the random variable $r$ will be considered as follows.

In the \textquotedblleft annealed\textquotedblright approximation, the fluctuation of the random variable $r$ is regarded as being fast such that its value at equation (\ref{eq:master-random-network}) can be replaced by its mean value, $\langle r\rangle$, where the mean is taken with respect to the Bernoulli distribution
\begin{eqnarray}
\mathcal{P}(r) = \left(1-c\right)\delta_{r,0} + c\delta_{r,1}\,,
\label{Pr}
\end{eqnarray}
which resembles (\ref{eq:bernoulli}) except for the fact that here $r_{s}=r$ for any link. Consequently, in the annealed case, one should examine the master equation
\begin{eqnarray}
p_{r}(k,s,t+1) = r\Pi(k-1, t)p_{r}(k-1,s,t) + \left[1-\Pi(k, t)\right]p_{r}(k,s,t)\,,
\label{me_r}
\end{eqnarray}
where one replaces the variable $r$ with the number $\langle r\rangle=c$, according to (\ref{Pr}). The equation can be solved by standard techniques \cite{dorogovtsev2003evolution} to obtain $p_{\langle r\rangle}(k, s, t)$ and, then, the reduced degree distribution $\tilde{P}_{\langle r\rangle}(k)$.

On the other hand, in the \textquotedblleft quenched\textquotedblright limit, the variable $r$ does not change during the analysis, which means that one requires the solution of the master equation (\ref{me_r}) with $r$ fixed, and the average $\langle p_{r}(k, s, t)\rangle$ is taken at the end in order to determine the stationary quantity $\langle\tilde P_{r}(k)\rangle$.

We will now analyze two dynamics for the growing network. In the first case, all of the pre-existing node, to which the new incoming vertex links, have the same probability to be connected. In the second dynamics, the new incoming vertex attaches to another according to the linear preferential linking.

\subsection{Random network with equiprobable attachment}

\begin{figure}[H]
  \centering
    \includegraphics[width=0.8\textwidth,height=\textheight,keepaspectratio]{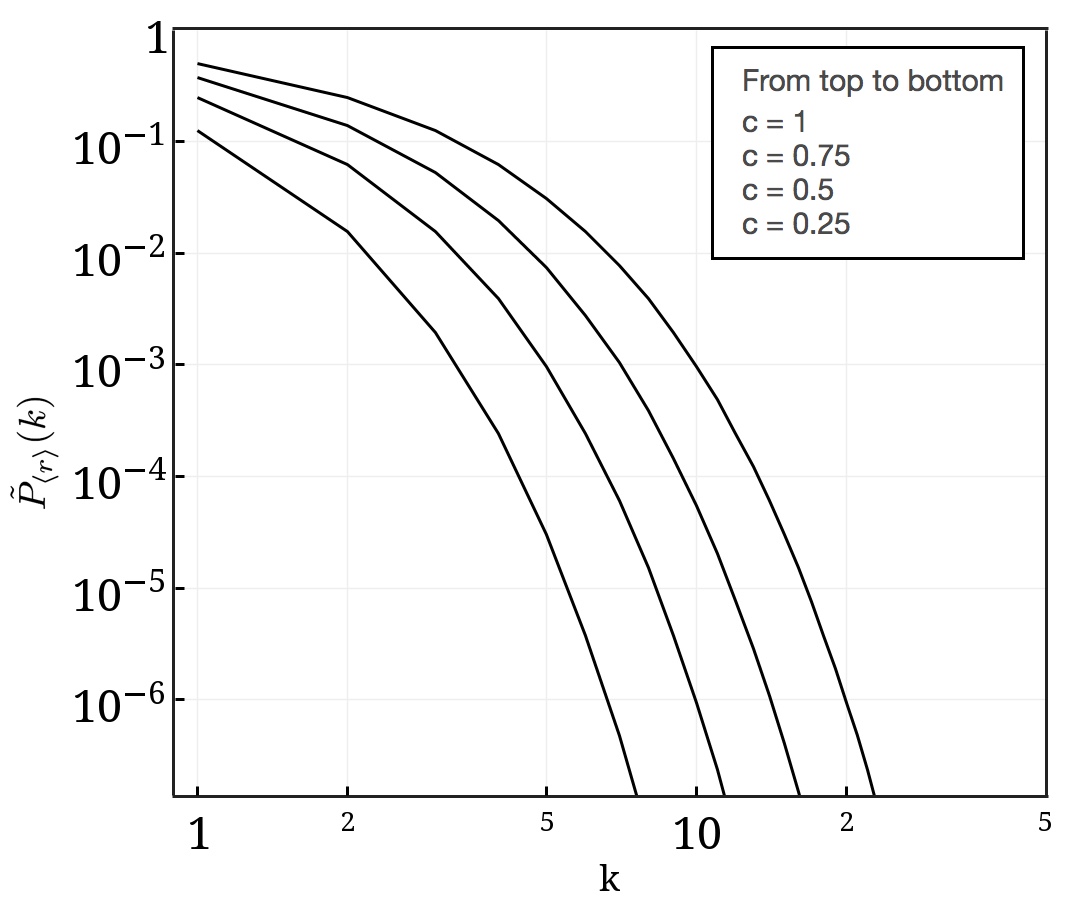}
    \caption[Distribution (log-log) of epidemic network using annealed approach]{Reduced distribution of epidemic network using annealed approach over a random network with equiprobable attachment, where $c=\{1,.75,.5,.25\}$}
  \label{fig:annealed_approach_random}
\end{figure}

\begin{figure}[H]
  \centering
    \includegraphics[width=0.8\textwidth,height=\textheight,keepaspectratio]{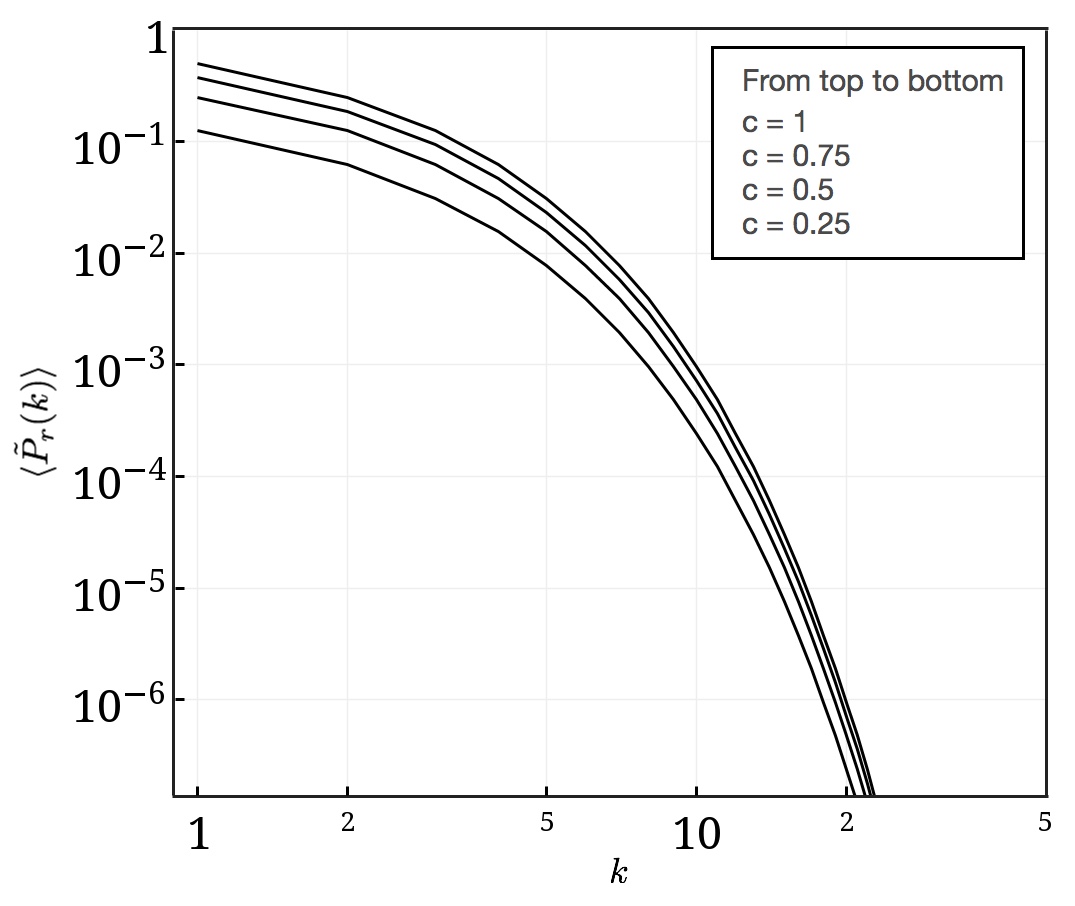} 
    \caption[Distribution (log-log) of epidemic network using quenched approach]{Reduced distribution of epidemic network using quenched approach over a random network with equiprobable attachment, where $c=\{1,.75,.5,.25\}$.}
  \label{fig:quenched_approach_random}
\end{figure}

In the present case, the probability of linking at time $t$ is $\Pi(k, t)=1/t$, and the solution of master equation can be obtained by usual techniques \cite{dorogovtsev2003evolution} to yield
\begin{eqnarray}
 \tilde{P}_{r}(k) = \left(\frac{r}{2}\right)^{k}\,.
\end{eqnarray}
Therefore, in the annealed case, one has
\begin{eqnarray}
 \tilde{P}_{\langle r\rangle}(k) = c^{k} 2^{-k} \qquad\textnormal{(annealed)}\,,
\end{eqnarray}
which is showed in Figure \ref{fig:annealed_approach_random}, while in the quenched case, the reduced distribution is
\begin{eqnarray}
\label{eq:quenched_approach_random}
\langle \tilde{P}_{r}(k)\rangle = c2^{-k} \qquad\textnormal{ (quenched)}\,,
\end{eqnarray}
and the reduced distribution can be seen in Figure \ref{fig:quenched_approach_random}.

One can see that while in the quenched case the reduced distribution differs only by a factor from the degree distribution, in the annealed case the difference is more significant in the sense that the basis of the exponent changed. For a given degree $k>1$, one can also see that $\tilde P_{\langle r\rangle}(k)<\langle\tilde P_{r}(k)\rangle$, showing that the quenched variable for contagion is more efficient than the annealed one in this model.

\subsection{Preferential attachment network}

\begin{figure}[H]
  \centering
    \includegraphics[width=0.8\textwidth,height=\textheight,keepaspectratio]{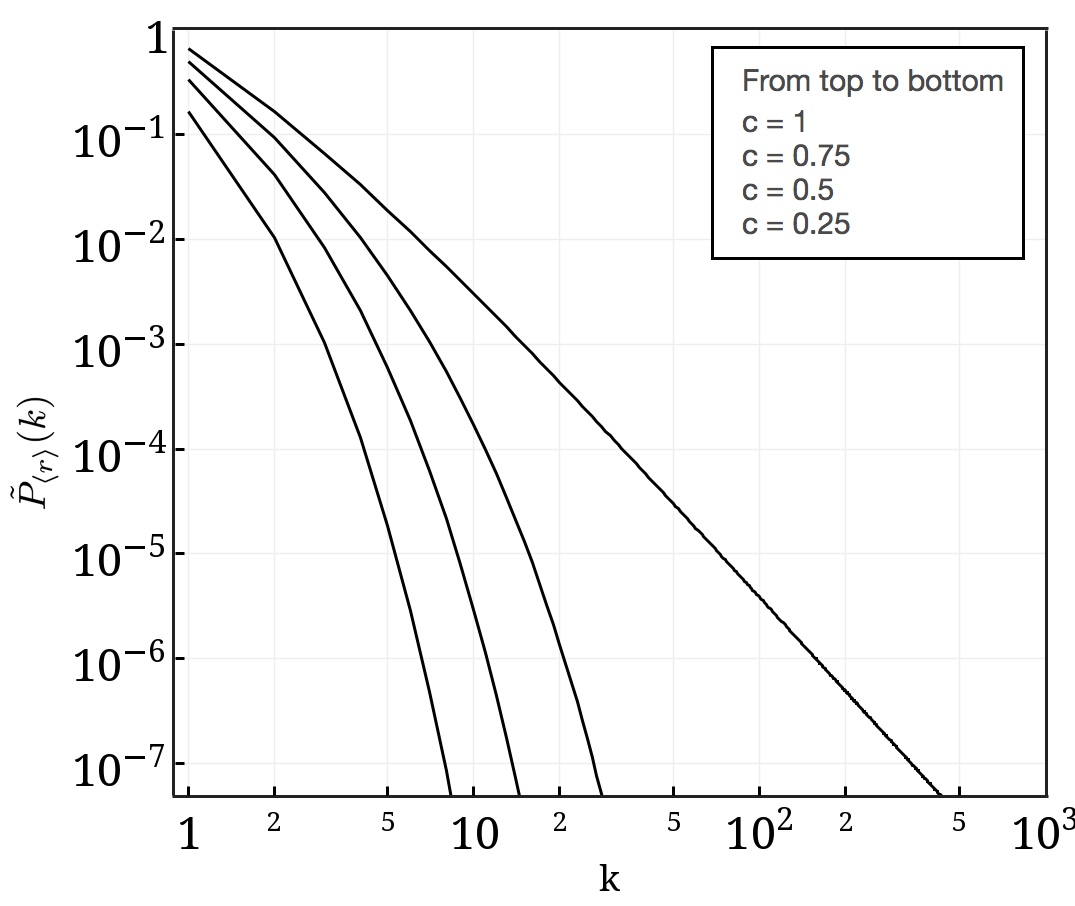}
  \caption[Distribution (log-log) of epidemic network using annealed approach]{Reduced distribution of epidemic network using annealed approach over a preferential attachment network, where $c=\{1,.75,.5,.25\}$}
  \label{fig:annealed_approach_preferential}
\end{figure}

\begin{figure}[H]
  \centering
    \includegraphics[width=0.8\textwidth,height=\textheight,keepaspectratio]{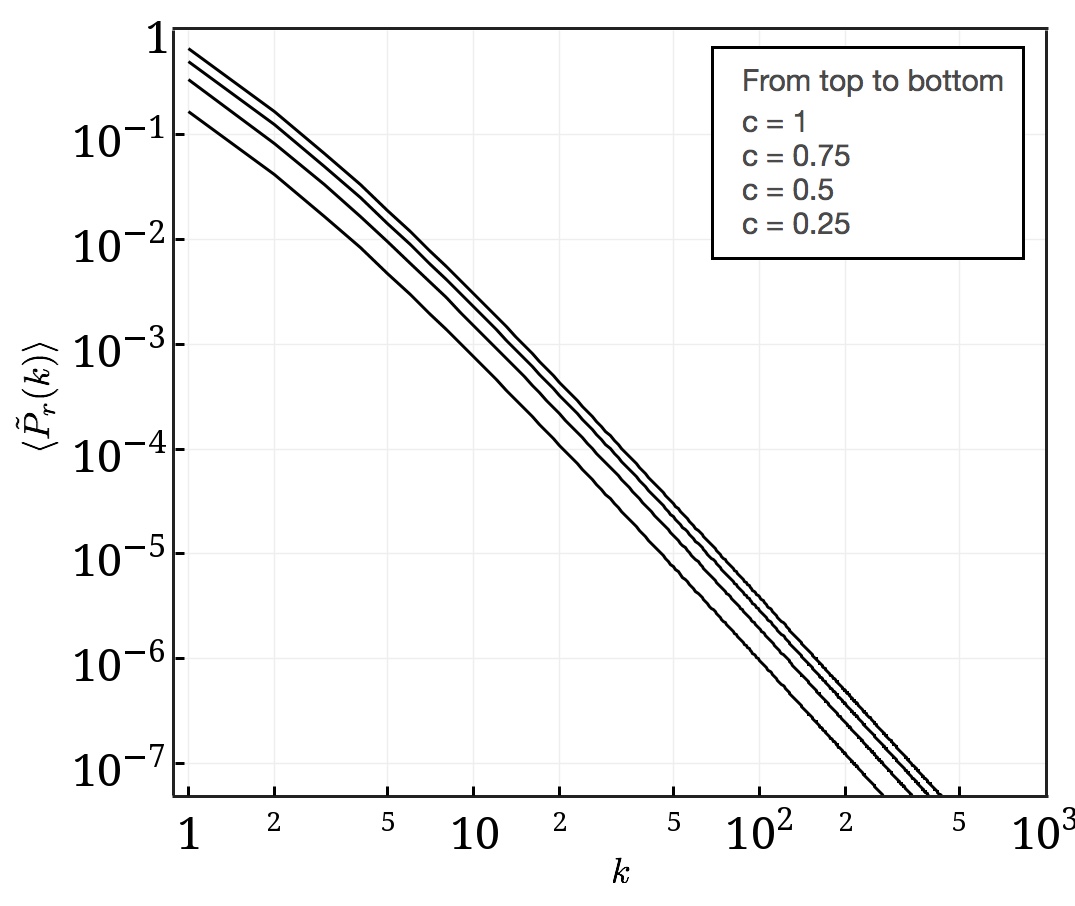}
  \caption[Distribution (log-log) of epidemic network using quenched approach]{Reduced distribution of epidemic network using quenched approach over a preferential attachment network, where $c=\{1,.75,.5,.25\}$}
  \label{fig:quenched_approach_preferential}
\end{figure}

In this section, the linking probability is $\Pi(k, t)=\frac{k}{2t}$ according to the model introduced before, where the denominator $2t$ is the normalization factor of the number of degrees. Solving the master equation in this case leads to
\begin{eqnarray} \label{eq:annealed_approach_preferential}
\tilde{P}_{\langle r\rangle}(k)  = \frac{4c^{k}}{k\left(k+1\right)\left(k+2\right)},
\end{eqnarray}
in the annealed situation (illustrated in Figure \ref{fig:annealed_approach_preferential}), while one has
\begin{eqnarray} \label{eq:quenched_approach_preferential}
\langle \tilde{P}_{r}(k)  \rangle = \frac{4c}{k\left(k+1\right)\left(k+2\right)},
\end{eqnarray}
in the quenched case, showed in Figure \ref{fig:quenched_approach_preferential}

Note that the effect of annealed variable has major consequences in this case: the reduced degree distribution is not a power-law function anymore. This means that the subgraph of contagion has a different structure from the underlying network. On the other hand, the quenched variable still preserves the power-law structure. As in the previous case (equiprobable attachment model), $\tilde P_{\langle r\rangle}(k)<\langle\tilde P_{r}(k)\rangle$ for $c\neq 1$ and $k>1$.

\section{Link-dependent contagion process}
\label{contagion}

In this section, the master equation
\begin{eqnarray}
p_{r}(k, s, t+1) = r_{t}\Pi(k-1, t)p_{r}(k-1, s, t) + \Big[ 1-\Pi(k, t) \Big]p_{r}(k, s, t)
\label{me}
\end{eqnarray}
will be analyzed when the underlying network follows a preferential linking prescription and the contagion linking, governed by the random variable $r_{s}\in\{0,1\}$ ($1\leq s\leq t$), depends on the particular edge considered. From the structure of the equation, it is clear that the annealed approximation leads to the same results as before, and one needs to analyze the quenched case only. In order to determine the reduced distribution $\langle \tilde{P}_{r}(k)\rangle$, one should first evaluate 
\begin{eqnarray}
\langle p_{r}(k, s, t)\rangle := \int dr_{1}\cdots dr_{t}\mathcal{P}(r_{1},\ldots,r_{t})p_{r}(k,s,t)\,.
\label{<pr>}
\end{eqnarray}
Since the random variables $\{r_{s}\}$ are independent and identically distributed, the global distribution can be factorized as $\mathcal{P}(r_{1},\ldots,r_{t})=\prod_{s=1}^{t}\mathcal{P}(r_{s})$, where
\begin{eqnarray}
\mathcal{P}(r_{s}) = \left(1-c\right)\delta_{r_{s},0} + c\delta_{r_{s},1}\,
\label{bernoulli}
\end{eqnarray}
as in (\ref{eq:bernoulli}), and, fixing the site $t$, one has
\begin{eqnarray}
\langle p_{r}(k, s, t)\rangle := \int dr_{t}\mathcal{P}(r_{t})p_{r_{t}}(k,s,t)\,,
\end{eqnarray}
where the $p_{r_{t}}$ stands for the marginal distribution
\begin{eqnarray}
p_{r_{t}}(k,s,t) = \int dr_{1} \cdots dr_{t-1}p_{r}(k,s,t).
\end{eqnarray}
Furthermore, from the obvious equality
\begin{eqnarray}
\fl p_{r_{t}}(k,s,t) = \left(1-r_{t}\right)p_{r_{t}}(k,s,t) + q_{r_{t}}(k,s,t)\,,\;\textnormal{ where }\; q_{r_{t}}(k,s,t) := r_{t}p_{r_{t}}(k,s,t)\,,
\end{eqnarray}
the desired quantity is evaluated as
\numparts \begin{eqnarray}
\fl \langle p_{r}(k, s, t)\rangle & = \int_{0}^{1}dr_{t}\mathcal{P}(r_{t})p_{r_{t}}(k,s,t) \\ 
& = \int_{0}^{1}dr_{t}\mathcal{P}(r_{t}) \left(1-r_{t}\right)p_{r_{t}}(k,s,t) + \int_{0}^{1}dr_{t}\mathcal{P}(r_{t}) q_{r_{t}}(k,s,t)\,.
\label{int01}
\end{eqnarray} \endnumparts

Nevertheless, instead of equation (\ref{int01}), let us consider
\begin{eqnarray}
\fl \int_{0}^{R}dr_{t}\mathcal{P}(r_{t})p_{r_{t}}(k,s,t) = & \int_{0}^{R}dr_{t}\mathcal{P}(r_{t}) \left(R-r_{t}\right)p_{r_{t}}(k,s,t)\,+ \int_{0}^{R}dr_{t}\mathcal{P}(r_{t}) q_{r_{t}}(k,s,t)\,.\label{trick}
\end{eqnarray}

It is evident that (\ref{trick}) recovers (\ref{int01}) when $R=1$. By invoking the Laplace transform on (\ref{trick}), one has
\begin{eqnarray}
\psi^{L}(\zeta) = \left(\frac{\zeta}{\zeta-1}\right)\chi^{L}(\zeta)\,,
\label{psichi}
\end{eqnarray}
where $\psi^{L}$ and $\chi^{L}$ are Laplace transforms of the functions
\begin{eqnarray}
\psi(R) := \int_{0}^{R}dr_{t}\mathcal{P}(r_{t})p_{r_{t}} \quad\textnormal{ and }\quad \chi(R) := \int_{0}^{R}dr_{t}\mathcal{P}(r_{t})q_{r_{t}}\,,
\end{eqnarray}
respectively. The function (\ref{<pr>}) is determined by first obtaining $\psi(R)$ and letting $R\rightarrow 1$. One may solve the equation (\ref{psichi}) and find that
\begin{eqnarray}
\psi(R) = \int_{0}^{\infty}dr_{t}\mathcal{P}(r_{t})q_{r_{t}}e^{R-r_{t}}\theta(R-r_{t})\,,
\label{psi}
\end{eqnarray}
where $\theta$ stands for the Heaviside function with $\theta(0)=1$.

\begin{figure}[H]
  \centering
    \includegraphics[width=0.8\textwidth,height=\textheight,keepaspectratio]{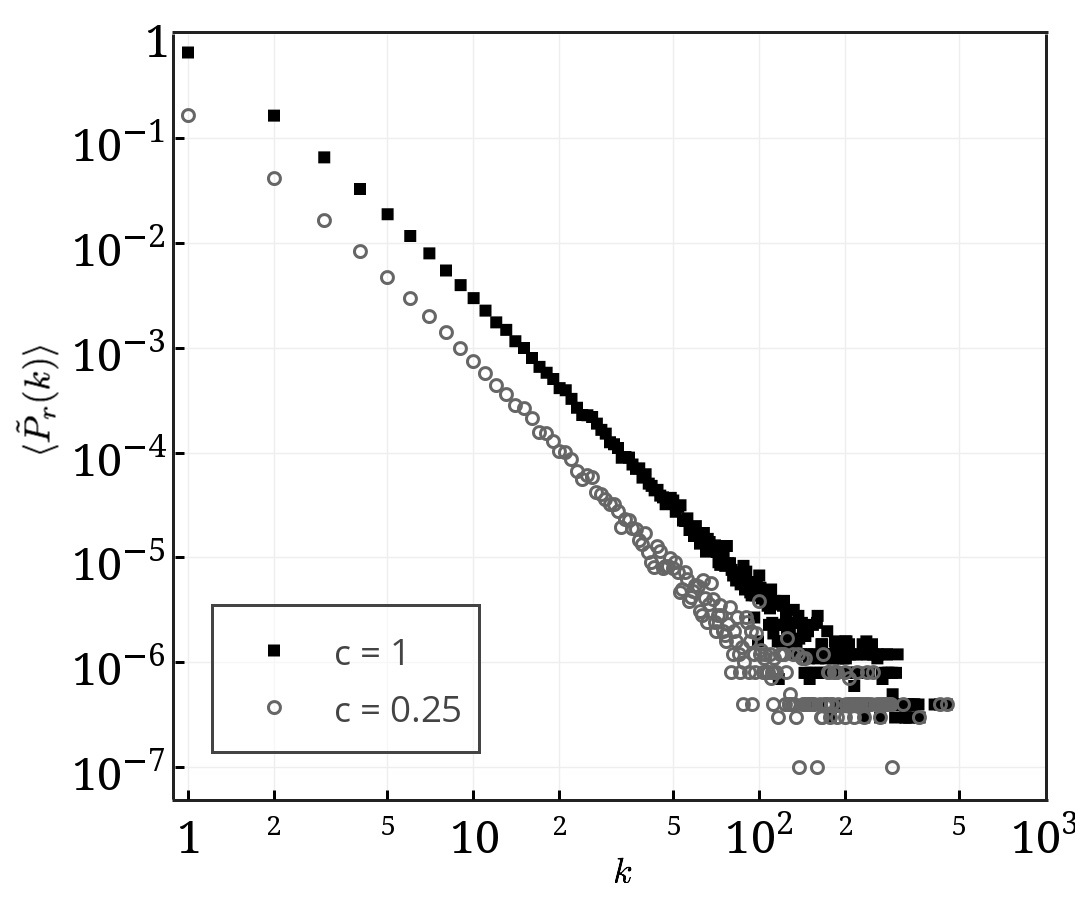}
  \caption[Simulation]{Reduced distribution obtained by simulation of networks generated by preferential attachment rule and the contagion variable is quenched. The curves decay as $k^{-3}$.}
  \label{fig:sim-quenched-fixed-preferential}
\end{figure}

In order to continue the calculations, one needs to evaluate the function $q_{r_{t}}$ above, and this can be done from the master equation (\ref{me}). Taking advantage of the fact that the random variable $r_{t}$ is binary (note that $r_{t}=r_{t}^{2}$), multiplying the master equation by $r_{t}$ and integrating in order to analyze the marginal distribution $p_{r_{t}}$ yields
\begin{eqnarray}
q_{r_{t}}(k, s, t+1) = \Pi(k-1, t)q_{r_{t}}(k-1, s, t) + \Big[ 1-\Pi(k, t) \Big]q_{r_{t}}(k, s, t)\,.
\label{meq}
\end{eqnarray}
If the underlying graph obeys the linear preferential linking rule, $\Pi(k, t)=k/2t$, and from the Z-transform
\begin{eqnarray}
Q_{r_{t}}^{Z}(z,s,t) = \sum_{k=1}^{\infty}z^{k}q_{r_{t}}(k,s,t)\,,
\end{eqnarray}
the recurrence equation (\ref{meq}) can be casted as
\begin{eqnarray}
\frac{\partial}{\partial t}Q_{r_{t}}^{Z}(z,s,t) + \frac{z\left(1-z\right)}{2t}\frac{\partial}{\partial z}Q_{r_{t}}^{Z}(z,s,t)=0
\end{eqnarray}
with the boundary condition $Q_{r_{t}}^{Z}(z,s,t=s) = r_{t}z$ for large times. This equation can be solved by the methods of characteristics and inverting the Z-transform, one has
\begin{eqnarray}
q_{r_{t}}(k,s,t) \sim \frac{r_{t}}{(\frac{t}{s})^{1/2}-1}\left(1-\left(\frac{s}{t}\right)^{1/2}\right)^{k}\qquad (t\gg 1)\,.
\label{qrs}
\end{eqnarray}
Finally, inserting (\ref{bernoulli}) and (\ref{qrs}) into (\ref{psi}), and letting $R\rightarrow 1$, one has
\begin{eqnarray}
\langle p_{r}(k,s,t)\rangle = \frac{c}{(\frac{t}{s})^{1/2}-1}\left(1-\left(\frac{s}{t}\right)^{1/2}\right)^{k}
\end{eqnarray}
for $t\gg 1$. We have then the reduced degree distribution, which is a stationary quantity, by the procedure described in (\ref{eq:Pred}). It turns out that the reduced distribution is just a constant multiplying the original one,
\begin{eqnarray}
\langle  \tilde{P}_{r_{s}}(k)\rangle = \frac{4c}{k\left(k+1\right)\left(k+2\right)}\,.
\label{Prs_BA}
\end{eqnarray}
As a way to support this analytical result, we performed a simulation of this model, where we generated $10^{3}$ networks with $10^{4}$ nodes each one (figure \ref{fig:sim-quenched-fixed-preferential}). The uncertainty bars in the figure are smaller than the size of the points. Works like \cite{Saltelli2007} study with detail the uncertainty in the output by relating it with the model input, which can be done in many different contexts \cite{Convertino2014,Ludtke2008}.

Following the same procedure, one can also show that
\begin{eqnarray}
\label{eq:dynamic_approach_random}
\langle  \tilde{P}_{r_{s}}(k)\rangle = c2^{-k}
\end{eqnarray}
if the underlying graph is made by uniform random linking, $\Pi(k,t)=1/t$.

Note that the results (\ref{Prs_BA}) and (\ref{eq:dynamic_approach_random}) are the same as (\ref{eq:quenched_approach_random}) and (\ref{eq:quenched_approach_preferential}), respectively. Although the reduced distributions are the same, the possible configurations of the contagion network are clearly different.

\section{Discussion}
\label{discussion}

In this work, we have shown an analysis based on analytical results supported by some simulations. Firstly, the quenched approach has better performance to spread an epidemy in all scenarios. It means, in simple words, that \textquotedblleft it is better to share, spread or infect \textit{sometimes}  a \textit{completely} object, information or disease, than propagating \textit{all the time}, but with just \textit{a part} of the information, disease or any element to spread\textquotedblright. We should also note that the present scheme of defining propagation could capture distinct behaviors in the contagion and its underlying graph: even the connections between vertices following a power-law distribution, the subgraph of contagion process may not follow the same distribution, as shown in the annealed case (see Figure \ref{fig:annealed_approach_preferential}). Nevertheless, one should remember that in the present setup, the problem of propagation of information over the network is associated to the presence of contagion links only. In a system where the vertices are spins (therefore, they have states that can influence other vertices) and the links can be favorable/unfavorable to align them, constitutes a different way to deal with the contagion problem, which can be mapped to a spin glass problem. In this case, one can see that an spreading due to an annealed random contagion variable is more efficient than the quenched one \cite{Mendes2008}.

Another important result is the reduced degree distribution between link-variable (section \ref{contagion}) and fixed contagion (section \ref{fixed}) being the same. This coincidence can be associated to the distribution of contagion variable being an independent and identically distributed random variable and absence of interaction between vertices. As a consequence, a single variable that decides the contagion of the entire graph (fixed contagion case) and the situation where each one of the links obey the same distribution independently have similarities, despite the fact that the configuration of infected links are clearly different in both cases.

\ack

M.C. acknowledges OEA scholarship for financial support.

\bibliography{references}

\end{document}